# Influence of Antireflection Si coatings on the Damage Threshold of fused silica upon irradiation with Mid-IR femtosecond laser pulses

## GEORGE  D. TSIBIDIS[1,a], AND EMMANUEL STRATAKIS[1,2b]

[1] *Institute of Electronic Structure and Laser (IESL), Foundation for Research and Technology (FORTH), Vassilika Vouton, 70013, Heraklion, Crete, Greece*
[2] *Department of Physics, University of Crete, 71003, Heraklion, Greece*

[a] *e-mail: tsibidis@iesl.forth.gr,* [b] *e-mail: stratak@iesl.forth.gr*

---

**Abstract:**

**Recent progress in the development of high-power mid-IR laser sources and the exciting laser driven physical phenomena associated with the irradiation of solids via ultrashort laser pulses in that spectral region are aimed to potentially create novel capabilities for material processing. In particularly, the investigation of the underlying physical processes and the evaluation of the optical breakdown threshold (OBT) following irradiation of bulk dielectric materials with Mid-IR femtosecond (fs) pulses has been recently presented. In this report, we will explore the conditions that generate sufficient carrier excitation levels which leads to damage upon irradiated a dielectric material ($SiO_2$) coated with antireflection (AR) semiconducting films (Si) of variable thickness with fs pulses. Simulation results demonstrate that the reflectivity and transmissivity of the $Si/SiO_2$ are thickness-dependent which can be employed to modulate the damage threshold of the substrate. The study is to provide innovative routes for selecting material sizes that can be used for antireflection coatings and applications in the Mid-IR region.**

Despite the extensive employment of ultrashort pulsed lasers for material processing, most industrial or technological applications have been based on the use of laser pulses between visible and near-IR wavelengths (see [1] and references therein). Various types of surface nano/micro-structured topographies have been produced by exploiting the wealth of possibilities femtosecond (fs) pulses in those spectral regions offers through modulation of the laser parameters.

The elucidation of the interaction of mid-IR fs pulses with fused silica ($SiO_2$) has provided crucial information for the behaviour of optical elements during experiments with intense mid-IR femtosecond pulses [2-6]. Simulation results validated by experimental data have confirmed the significant influence of the laser parameters on the damage threshold as well as the enhanced domination of tunnelling processes compared to multiphoton excitation [7]. The latter is attributed to the correlation of the Keldysh parameter $\gamma \sim 1/(\sqrt{I}\lambda_L)$ (where $I$ and $\lambda_L$ stand for the laser intensity and wavelength, respectively) with the wavelength [2] leading to values smaller than one.

Despite the interesting physical phenomena produced after irradiation of bulk solids with laser pulses at this spectral region [2], there exist fundamental open questions in regard to the effects of laser exposure of materials coated with semiconductors which are known to be transparent at Mid-IR. We recall that infrared antireflection coatings have been widely used in various applications (i.e. passive optical components) where the coating is applied to a low index substrate such as glass [8,9]. Similarly, nonlinear crystals coated with antireflection (AR) films have been proposed to generate coherent radiation in the Mid-IR region [8,10]. It is well known that AR coatings are applied to optical surfaces to increase the throughput of a system and reduce undesired effects induced by reflections. Although the influence of the coating on the damage threshold of the dielectric substrate is of paramount importance, a key challenge is the optimization of the features of the semiconducting layer (i.e. type of semiconductor, thickness) to exhibit maximum antireflection properties while achieving minimum damage on the dielectric material. Thus, a detailed investigation of the response of a two layered material is expected to provide the route for the fabrication of efficient optical components for Mid-IR experiments considering, also, the transparency of the semiconductor at long wavelengths.

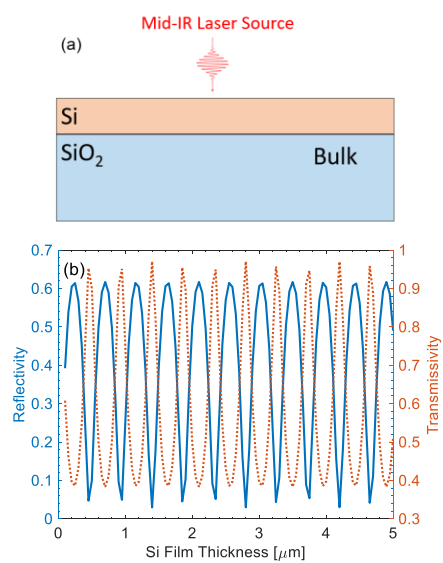

**Fig.1:** (a) Double layered material (Silicon films on top of bulk fused silica irradiated with a Mid-IR laser pulse), (b) Optical parameters of $Si/SiO_2$ as a function of Si thickness (at room temperature).

To address the above challenge, we will consider a two-layered material, Silicon/Fused Silica (hereinafter Si/SiO₂) (Fig.1a) that is irradiated with laser pulses with wavelength $\lambda_L$=3.2 μm and pulse duration $\tau_p$=170 fs. The selection of Silicon as a coating was based on the fact that a suitable material was required to have *high* transparency and *low* absorption coefficient at this spectral region; on the other hand, SiO₂ is a material that is conventionally used in optical components. Nevertheless, an important issue is whether the thickness of Si influences the laser energy which is transmitted into the bulk SiO₂ and therefore, the damage threshold of the substrate. In previous reports, experimental [11] and theoretical [12] studies demonstrated that the optical properties of the coating vary with its thickness while such a dependence disappears when the size of the film becomes significantly larger than the optical penetration depth. Hence, considering transparency of Si at $\lambda_L$=3.2 μm one could argue that both the amounts of the reflected energy (from the Si/SiO₂) and the transmitted (into SiO₂) are independent on the Si thickness.

An analysis is performed, firstly, on the optical properties of Si/SiO₂ for various thicknesses $d$ of the coating ($d$=0.1-5 μm). The reflectivity $R$, transmissivity $T$ and the absorbance $A=1-R-T$ are derived via the employment of the multiple reflection theory [13]. Hence, the following expressions are used to calculate the optical parameters for a thin film on a substrate (for a *p*-polarised beam) [12,13]

$$R = \mid r_{dl} \mid^2, \quad T = \mid t_{dl} \mid^2 \widetilde{N}_S, \quad r_{dl} = \frac{r_{am} + r_{mS} e^{2\beta j}}{1 + r_{am} r_{mS} e^{2\beta j}},$$

$$t_{dl} = \frac{t_{am} t_{mS} e^{\beta j}}{1 + r_{am} r_{mS} e^{2\beta j}}, \quad \beta = 2\pi d \widetilde{N}_m / \lambda_L \tag{1}$$

$$r_{am} = \frac{\widetilde{N}_m - \widetilde{N}_a}{\widetilde{N}_m + \widetilde{N}_a}, \quad r_{mS} = \frac{\widetilde{N}_S - \widetilde{N}_m}{\widetilde{N}_S + \widetilde{N}_m}, \quad t_{am} = \frac{2\widetilde{N}_a}{\widetilde{N}_m + \widetilde{N}_a}, \quad t_{mS} = \frac{2\widetilde{N}_m}{\widetilde{N}_S + \widetilde{N}_m}$$

where the indices '*a*', '*m*', '*S*' stand for 'air', 'Si', 'SiO₂', respectively. It is noted that the refractive indices of the materials such as air, Si and SiO₂ are equal to $\widetilde{N}_a = 1, \widetilde{N}_m = 3.4309$ [14], $\widetilde{N}_S = 1.4143$ [15], respectively at $\lambda_L$ = 3.2 μm (at room temperature and before irradiation). Simulation results indicate a dependence of $R$ and $T$ on $d$. While for all thicknesses $R+T$=1 which confirms that no energy is absorbed from Si, calculations show a periodic variation of the optical parameter values (see Supplementary Material). Interestingly, the values of the reflectivity are in the range between 0.03 to 0.62 while the amount of the energy transmitted into the substrate is between 0.38 and 0.97 of the deposited energy (Fig.1b). These predictions manifest that the (anti)reflection level of the coating can be controlled through an appropriate selection of the thickness of the film. It is evident that if other types of semiconductors are used as a coating [8], both the optical parameters and the value of $d$ for minimum reflectivity are expected to change, however, such an investigation is beyond the scope of the current study. It is emphasised that, while a part of the laser energy is absorbed from Si at lower wavelengths (see Supplementary Material), there is no energy absorption in Si film for Mid-IR pulses.

Apart from the enhancement of antireflection through the choice of $d$, to explore any damage conditions induced on the bi-layered material is of high importance to investigate the ultrafast dynamics following irradiation with fs Mid-IR pulses. As indicated in the previous paragraph, a variation of reflectivity with $d$ leads to a change in the amount of energy that will be transmitted (and eventually, absorbed) into SiO₂ affecting the excited electron dynamics in the substrate.

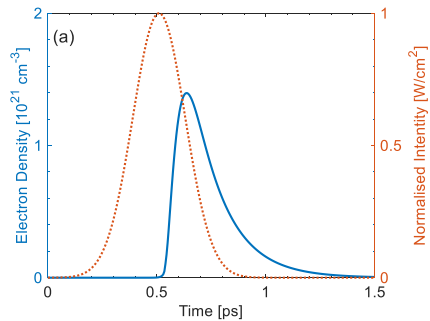

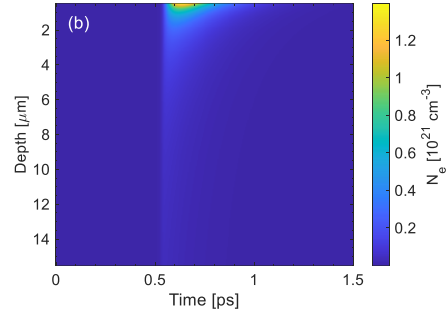

**Fig.2:** (a) Electron density evolution on the surface of SiO$_2$. (b) Electron density evolution as a function of time and depth (inside SiO$_2$) ($F$=3 J/cm$^2$, $d$=500 nm).

In the present work, the free electron generation in dielectrics following excitation with ultrashort pulsed lasers is described through the following two single rate equations (Eq.2); these equations provide both the evolution of the excited electron and self-trapped exciton (STE) densities, $N_e$ and $N_{STE}$ [3,16], respectively,

$$\frac{dN_e}{dt} = \frac{N_V - N_e}{N_V}\left(W_{PI}^{(1)} + N_e A^{(1)}\right) + \frac{N_{STE}}{N_V}\left(W_{PI}^{(2)} + N_e A^{(2)}\right) - \frac{N_e}{\tau_{tr}}$$
$$\frac{dN_{STE}}{dt} = \frac{N_e}{\tau_{tr}} - \frac{N_{STE}}{N_V}\left(W_{PI}^{(2)} + N_e A^{(2)}\right)$$

(2)

where $N_V$=2.2×10$^{22}$ cm$^{-3}$ corresponds to the valence electron density. In Eqs.2, the STE states are assumed to be centres situated at an energy level below the conduction band (i.e. $E_G^{(2)}$=6 eV); on the other hand, for fused silica, the band gap between the valence (VB) and the conduction band (CB) is $E_G^{(1)}$=9 eV [16,17]. It is noted that $\tau_r$ ∼150 fs [18] stands for the trapping time of electrons in STE states. In the above framework, photoexcitation assumes photoionization ($W_{PI}^{(i)}$) and impact ionization processes (i.e. $A^{(i)}$ stands for the avalanche ionisation rate) that will lead to transition from VB to CB ($i$=1) and from the STE level to the CB ($i$=2).

Results in Fig.2 illustrate the dynamics of the electron density on the surface of the dielectric material (Fig.2a) and the temporal evolution of $N_e$ on SiO$_2$ *inside* the material (Fig.2b) assuming irradiation with fluence $F$=3 J/cm$^2$ for a particular value of the thickness of the Si coating ($d$=500 nm). A detailed numerical approach is presented in the Supplementary Material. Similar results are deduced at other $d$ and $F$ values with an expected different maximum electron density as the absorbed energy from SiO$_2$ varies with $d$. It is evident that fused silica absorbs significantly leading to high levels of excitation (∼1.4×10$^{21}$ cm$^{-3}$). Thus, in contrast to the transparent behaviour exhibited from Si, fused silica absorbs yielding a gradually increasing excitation of carriers despite a higher band gap than Si.

While the analysis of the parameters illustrated in Fig.1b focused entirely on the optical response of the bi-layered structure at room temperature, a more useful investigation should involve the fingerprint of the ultrafast dynamics on the temporal variation of $R$ and $T$. More specifically, it has been demonstrated [19] that irradiation of solids with ultrashort pulses leads to a dynamical change of the optical parameters which influence, eventually, the energy absorption from the material (i.e. $\tilde{N}_S$ in Eq.1 becomes a complex number). Thus, before evaluating the correlation of the minimum energy required to damage SiO$_2$ with the coating thickness via the employment of fs pulses, one should explore, firstly, the levels of reflected/transmitted amounts of the laser energy derived from the density of excited electrons values. To argue on the impact of the thickness on $R$ and $T$, simulations of the dynamics of these parameters (i.e. considering that $R+T$=1, in Si) have been conducted for various values of $d$. In Fig.3a, $R$ is illustrated for $d$=200 nm and $d$=500 nm that demonstrates a remarkable temporal variation of $R$ for the higher thickness; by contrast, there is an insignificant variation of $R$ for larger $d$. To illustrate the impact of the coating thickness on the reflectivity, the minimum and maximum values of $R(t)$ as a function of $d$ is depicted in Fig.3b.

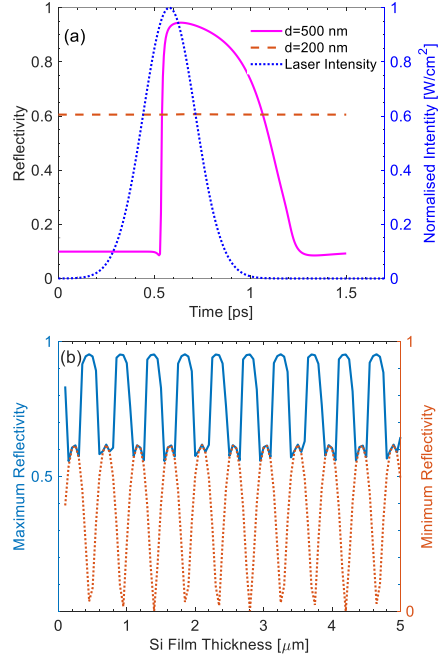

**Fig.3:** (a) Dynamics of $R$ for $d$=200 nm and $d$=500 nm, (b) Min-max values of reflectivity of Si/SiO₂ as a function of Si thickness ($F$=3 J/cm²).

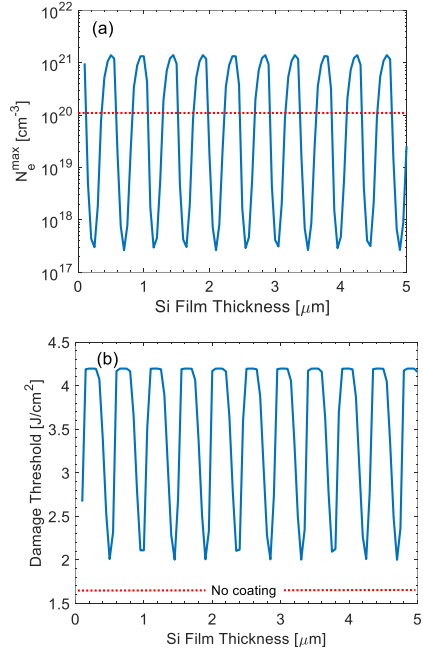

**Fig.4:** (a) Maximum electron density on the surface of SiO₂ at various $d$ ($F$=3 J/cm²) (*red* dotted line indicates OBT). (b) Damage threshold on SiO₂ surface as a function of $d$ (*red* dotted line indicates the damage threshold for bulk SiO₂ in absence of coating that occurs at $F$=1.65 J/cm² [2]).

On the other hand, to interpret the role of the initial reflectivity (Fig.1b) in the evolution of $R$, it is important to evaluate the dynamics of $N_e$. As the Si coating does not absorb any amount of the laser energy, the comparison of the transmitted portion of the energy into the dielectric and the evolution of the optical parameters as a result of the electron excitation on SiO₂ can explain the trend. More specifically, simulation results showed that for $d$=200 nm the initial transmissivity is $T(t$=0)=0.4 compared to $T(t$=0)=0.9 for $d$=500 nm. According to the theoretical predictions shown in Fig.2a, the absorbed energy from SiO₂ will generate excited electrons in the dielectric with a temporally varying density. Nevertheless, the difference in the absorbed energy for the two thicknesses yields significantly different excitation levels. In particular, results manifest that the maximum electron density $N_e^{max}$ on the surface of SiO₂ is $N_e^{max} \sim 4 \times 10^{17}$ cm⁻³ and $N_e^{max} \sim 1.4 \times 10^{21}$ cm⁻³ for $d$=200 nm and $d$=500 nm, respectively. The large excitation that occurs for $d$=500 nm in contrast to a very small production of excited electrons for $d$=200 nm influences the optical parameters of the substrate as emphasized above (and reported in previous works [19-21]). This impact is reflected on the optical properties of the complex Si/SiO$_2$; thus, the evolution of the reflectivity of a bi-layered structure with $d$=200 nm is significantly different from that with $d$=500 nm. It is important to note that the small initial reflectivity for $d$=500 nm (Fig.3a) leads to a large generation of $N_e$ that further increases the reflectivity; despite an expected, drop of the carrier density as a result of the enhanced reflectivity, the produced large $N_e$ is significantly high.

A parametric study has, also, been conducted to correlate the maximum carrier densities produced on the dielectric material due to electron excitation with the coating thickness (Fig.4a). Simulation results allow the estimation of the fluence threshold $F_{thr}$ at which the (maximum) value of $N_e$ exceeds a critical value $N_e^{cr}$ (i.e. $N_e^{cr} \equiv 4\pi^2 c^2 m_e \varepsilon_0/(\lambda_L^2 e^2)$) that is, usually, coined as the optical breakdown threshold (OBT) [18]; in the expression that gives $N_e^{cr}$, $c$ is the speed of light, $m_e$ stands for mass of electron, $e$ is the electron charge and $\varepsilon_0$ is the permittivity of vacuum). In particularly, $N_e^{cr} = 1.09 \times 10^{20}$ cm$^{-3}$ at $\lambda_L = 3.2$ μm. In previous reports, OBT has been associated with the damage threshold, although, a more accurate estimate is expected to be deduced via an energetic criterion (see Supplementary Material [18,19]). The capacity of specific values of the fluence to generate sufficient electron densities that excited SiO$_2$ above the OBT is illustrated in Fig.4a. More specifically, simulations have been conducted which show that the thickness of the coating influences the maximum carrier density $N_e^{max}$ on the surface of SiO$_2$ (the *red* dotted line in Fig.4a indicates the density of the produced electrons at $F$=3 J/cm$^2$).

Furthermore, a correlation of the damage threshold with the coating thickness has been produced and theoretical results are illustrated in Fig.4b. Predictions yield values for $F_{thr}$ in the range between 2.0 J/cm$^2$ and 4.2 J/cm$^2$. Calculations show that for all Si film thicknesses, the presence of the coating induces damage of SiO$_2$ at higher fluences than in the absence of the Si film. The latter occurs at ~1.65 J/cm$^2$ (*red* dashed line in Fig.4b) [2]. To interpret this trend, it is important to evaluate the role of the optical parameters and the level of the absorbed energy from the substrate. As demonstrated in this work (see also in Supplementary Material), the initial reflectivity appears to influence substantially the number of the generated carriers. For all Si thicknesses, the resulting reflectivity is always non-vanishing and larger than the reflectivity of SiO$_2$ which indicates that a higher fluence is required to damage the substrate compared to a bare SiO$_2$. To summarise the predictions illustrated in Fig.4b, simulations manifest that a modulation of the damage threshold is possible through an appropriate selection of the thickness of Si. The remarkably large window of fluence values that lead to damage in addition to the, also, achievable broad range of reflectivity values of the bi-layered material emphasise the critical role of the Si thickness.

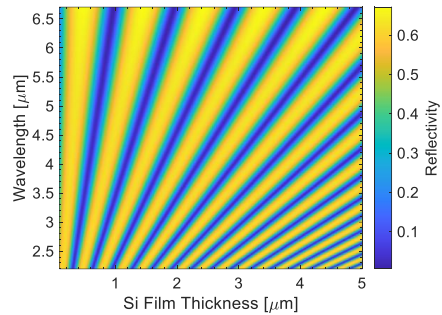

**Fig.5:** Reflectivity of Si/SiO$_2$ as a function of $d$ and $\lambda_L$.

The development of appropriate experimental protocols is certainly required to validate the theoretical model. In regard to the selection of Si as a coating, a more comprehensive, study of the performance of a range of semiconducting materials considering their optical, thermophysical and adhesive properties would allow to determine the optimal coating materials to meet the wavelength requirements. It is noted that although the focus of the current work is centred on the investigation of the response of the Si/SiO$_2$ at $\lambda_L$=3.2 μm, a similar exploration can be performed at other laser wavelengths (see Fig.5 and Supplementary Material). Due to the fact that the optical properties are dependent on the laser wavelength, more elaboration is required to deduce any enhancement of the antireflection level at various $\lambda_L$.

In conclusion, our approach aimed to present a thorough analysis of the influence of the thickness of semiconducting films placed on SiO$_2$ on both the antireflective levels of the complex and the damage induced on the dielectric substrate upon irradiation with fs pulses. Simulation results showed a remarkable impact of the thickness on these properties. The study is to provide innovative routes for material sizes that can be used for antireflection coating and applications in the Mid-IR region.

**Funding**. *NEP* project (GA 101007417); *HELLAS-CH* project (MIS 5002735).

**Disclosures.** The authors declare no conflicts of interest.

**Data availability.** Data underlying the results presented in this paper are not publicly available at this time but may be obtained from the authors upon reasonable request.

**Supplement Material.** See Supplement 1 for supporting content.

# Supplementary Material for

# Influence of Antireflection Si coatings on the Damage Threshold of fused silica upon irradiation with Mid-IR femtosecond laser pulses

## GEORGE  D. TSIBIDIS[1,a], AND EMMANUEL STRATAKIS[1,2b]

[1] *Institute of Electronic Structure and Laser (IESL), Foundation for Research and Technology (FORTH), Vassilika Vouton, 70013, Heraklion, Crete, Greece*
[2] *Department of Physics, University of Crete, 71003, Heraklion, Greece*

[a] *e-mail:* tsibidis@iesl.forth.gr, [b] *e-mail:* stratak@iesl.forth.gr

---

## 1. Damage threshold for Si/SiO₂

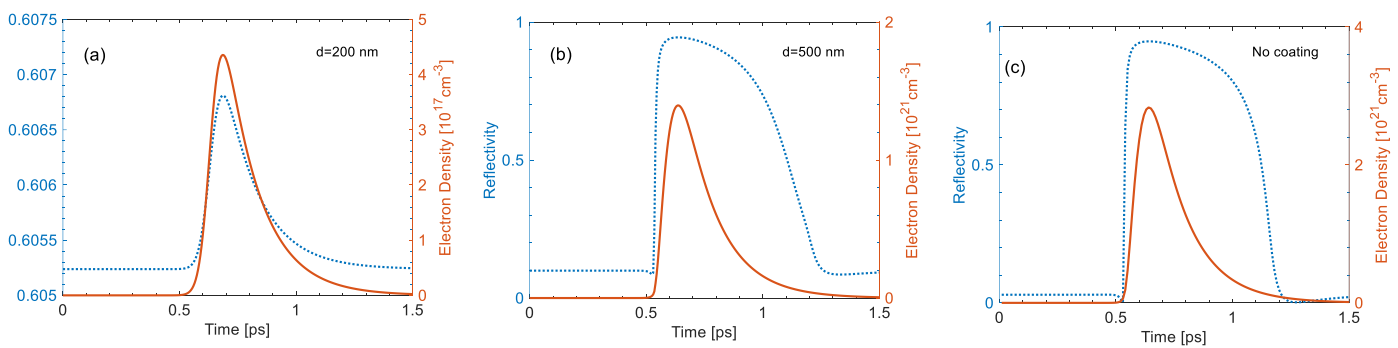

**Fig.SM1:** Reflectivity of Si/SiO₂ and electron density evolution on SiO₂ for (a) $d$=200 nm, (b) $d$=500 nm, (c) Absence of Si coating. (Simulations have been performed at 3 J/cm²).

Simulations illustrated in Fig.SM1 have been performed for three different thicknesses $d$ of the Si coating: (a) $d$=200 nm, (b) $d$=500 nm and (c) no coating (i.e. only SiO₂ is present). The selection of the $d$ values is to justify the behaviour depicted in Fig.5 (in the main manuscript) and interpret the behaviour of the damage threshold dependence as a function of the thickness. Similar conclusions can be deduced at other values of $d$. Simulations for (a) $d$=200 nm, (b) $d$=500 nm and (c) no coating yield maximum carrier densities $N_e^{max} \sim 4.4 \times 10^{17}$ cm$^{-3}$, $N_e^{max} \sim 1.4 \times 10^{21}$ cm$^{-3}$ and $N_e^{max} \sim 2.8 \times 10^{21}$ cm$^{-3}$, respectively. A comparison of the initial reflectivities for the two thicknesses and in absence of the Si coating indicates that the initial value plays a significant role in the energy absorption from the substrate despite an increase of the reflectivity at later stages. More specifically, at this fluence (3 J/cm$^2$), the increase of the carriers as the laser intensity rises leads to an increase of the reflectivity after a small initial drop (see also Fig.3a); nevertheless, the increase of the reflectivity is not sufficient to inhibit the steep rise of $N_e$. In Fig.SM2, the maximum electron density on the surface of SiO$_2$ at various $d$ is illustrated. It is shown that the enhanced reflectivity of the double-layered complex due to the presence of the Si coating reduces the amount of the absorbed energy and therefore, a smaller number of excited carriers is produced in contrast to the case in which Si coating is absent (*red* dashed-dotted line in Fig.SM2).

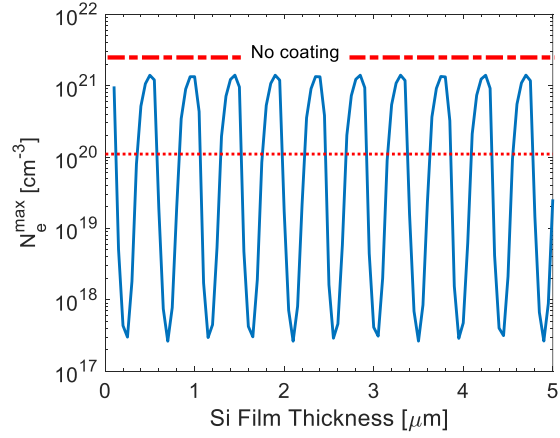

**Fig.SM2:** Maximum electron density on the surface of SiO$_2$ at various $d$ ($F$=3 J/cm$^2$) (*red* dotted line indicates OBT, *red* dashed-dotted line indicates the maximum electron density in absence of any coating).

## 2. Reflectivity for laser wavelengths in the range between 2.2 μm and 6.7 μm-Optical Kerr effect

To evaluate the impact of the thickness $d$ of Si and the laser wavelength $\lambda_L$ on the optical parameters, simulations have been performed through the employment of Eq.SM.1 assuming wavelengths in the range [2.2-6.7 μm]. The choice of the wavelengths was based on the availability of bibliographic data for the refractive index for Si [1] and SiO$_2$ [2]. More specifically, the following values for the refractive index are used

$$
\begin{aligned}
n^2 &= 11.67316 + \frac{1}{\lambda_L^2} + \frac{0.004482633}{\lambda_L^2 - 1.108205^2} && \text{(For Silicon)} \\
n^2 &= 1 + \frac{0.6961663\lambda_L^2}{\lambda_L^2 - 0.0684043^2} + \frac{0.4079426\lambda_L^2}{\lambda_L^2 - 0.1162414^2} + \frac{0.8974794\lambda_L^2}{\lambda_L^2 - 9.896161^2} && \text{(For Fused Silica)}
\end{aligned}
\tag{SM.1}
$$

where $\varepsilon = (n + ik)^2$. It is noted that $\varepsilon$ is the dielectric function for the unexcited material while $n$, $k$ are the real refractive index and the extinction coefficient, respectively. The dielectric function is dependent on the excitation level (i.e. the density of the excited carriers) and the material (the superscript $i$ characterizes the material, $i$=1 or 2 for Si or SiO$_2$, respectively).

$$
\varepsilon^{(i)} = 1 + \left(\varepsilon_{un}^{(i)} - 1\right)\left(1 - \frac{N_e^{(i)}}{N_V^{(i)}}\right) \frac{e^2 N_e^{(i)}}{m_r m_e \varepsilon_0 \omega_L^2} \frac{1}{\left(1 + i\frac{1}{\omega_L \tau_c}\right)}
\tag{SM.2}
$$

In Eq.SM.2, $\varepsilon_{un}^{(i)}$ stands for the laser wavelength dependent dielectric parameter of the unexcited material, $N_e^{(i)}$ and $N_V^{(i)}$ are the carrier densities in the conduction and valence bands ($N_V^{(1)} = 5 \times 10^{22}$ cm$^{-3}$ [3] and $N_V^{(2)} = 2.2 \times 10^{22}$ cm$^{-3}$ [4]), respectively, $m_e$ is the electron mass, $e$ is the electron charge, $m_r = 0.18$, $\varepsilon_0$ is the vacuum permittivity, $\omega_L$ is the laser frequency and $\tau_c = 5$ fs stands for the electron collision time (see Ref. [4] for more details on the selection of this value) .

In previous works, the contribution of optical Kerr effect to the refractive index of the irradiated material for Si and SiO$_2$ for mid-IR laser pulses was reported [3,4]. Kerr effect causes an intensity dependent variation of the refractive index of the material $n = n_0 + n_2 I$, where $n_0$ stands for the refractive index in absence of the Kerr effect, while $n_2$ is the second-order nonlinear refractive index. The variation of the refractive index due to Kerr effect and the impact on the ultrafast dynamics and thermal phenomena at various wavelengths were described in detail in Refs. [3,4] and it is not within the scope of the current study to provide a comparison of results within and without the contribution of the optical Kerr effect. Furthermore, values of $n_2$ for both materials were provided as a function of the laser wavelength for both materials. For the sake of completeness, all calculations presented in the main manuscript in the current work at $\lambda_L = 3.2$ μm have been performed assuming the presence of Kerr effect.

Simulations for optical parameters (at room temperature) show that at increasing wavelength the periodic variation of the reflectivity (Fig.SM3a-c) and transmissivity (Fig.SM3d) is characterized from a larger value of the periodicity. Interestingly (Fig.SM3b), results demonstrate that the reflectivity can drop to significantly small values by an appropriate selection of the wavelength and the Si thickness. Simulations are shown in Fig.SM3b for the reflectivity variation as a function of the thickness at three different wavelengths ($\lambda_L = 3.2$ μm, 5 μm and 6 μm). For the same thickness $d$, a spatial variation of the reflectivity occurs (see Fig.SM3c for $d$=2.8 μm). The periodicity at various wavelengths as a function of the thickness is explained in the next Section. Results manifest that appropriate combinations of the Si film thickness and laser wavelength can be used to achieve optimum (i.e application based) reflectivity/transmissivity values. The periodicity of the optical parameters as a function of the Si thickness is explained via the multireflection theory and the interference between the reflected and incident waves (see main manuscript and Section 3).

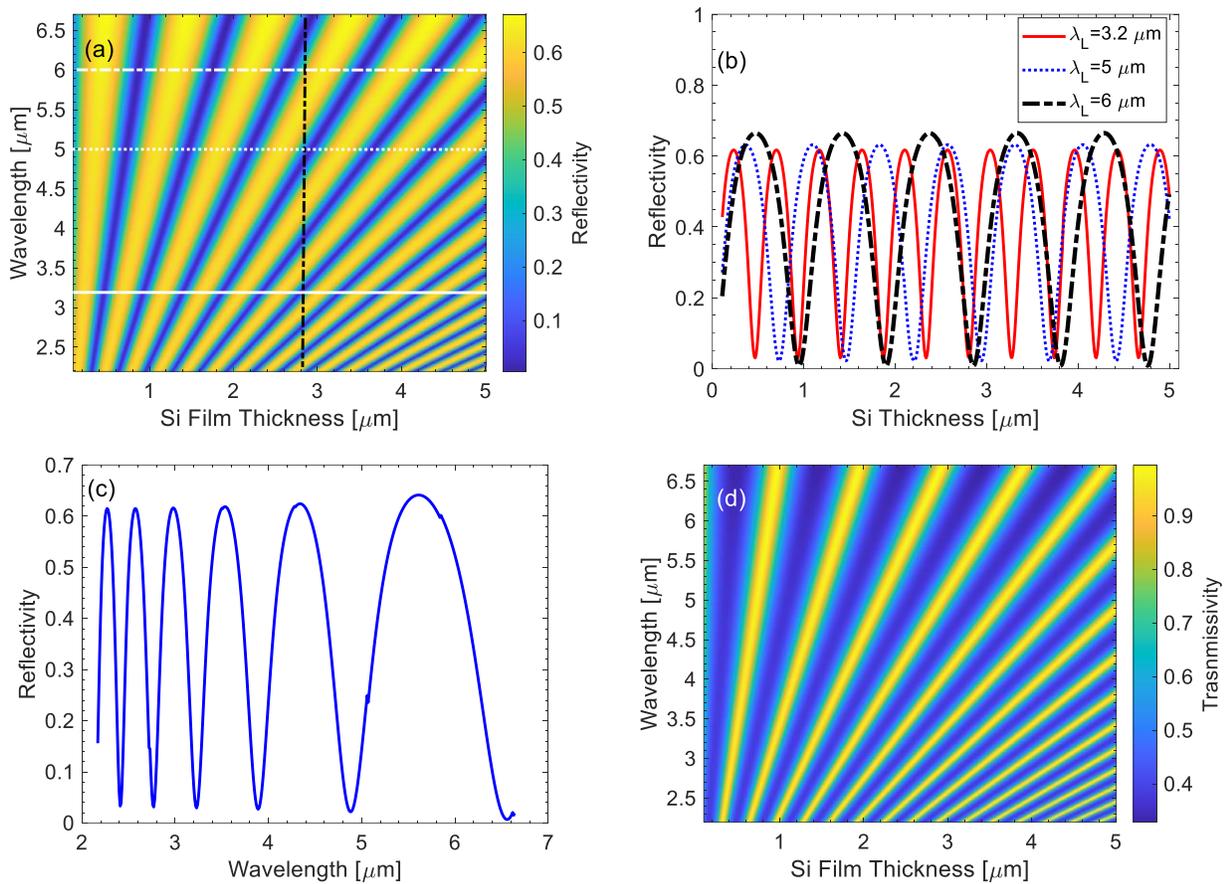

**Fig.SM3:** (a) Reflectivity of Si/SiO₂ as a function of $d$ and the laser wavelength, (b) Reflectivity of Si/SiO₂ as a function of $d$ along the *white* lines in (a) corresponding to three different wavelengths, (c) Reflectivity of Si/SiO₂ along the *black* dashed-dotted line in (a), at $d$=2.8 μm), (d) Transmissivity as a function of $d$ and the laser wavelength.

## 3. Periodicity of optical parameters at various thicknesses

The pronounced periodicity of the optical parameters and the carrier density dependence as a function of thickness demonstrated in Figs.2,4-6 (main manuscript) are due to the term $\beta = 2\pi d \tilde{N}_m / \lambda_l$ in the expression that provides the reflectivity and the transmissivity (see Eq.1 in main manuscript). As seen in Eq.1, $\beta$ is included in $e^{2\beta j}$ which leads to the (spatially) periodic behaviour of the optical parameters. Thus, the absorbed energy, excited carriers and damage threshold, in turn, are expected to follow a similar trend. By contrast, Fig.SM3a,d indicate that the periodicity at increasing wavelength increases. This is due to the fact that $\tilde{N}_m$ drops at increasing $\lambda_L$ (see expression for the refractive index of fused silica in Section 2) which yields a smaller value of $\beta$ and therefore, a larger periodicity (Fig.SM3a). To illustrate this decrease, $\beta$ has been plotted at increasing wavelength (for example, at $d=2$ μm as shown) (Fig.SM3b). It is noted that this periodic dependence of the optical parameters on the thickness of the material (Si) that lies on top of SiO$_2$ is a characteristic feature of the behaviour of Si at Mid-IR (i.e. zero absorption). By contrast, at lower laser wavelengths for which absorption levels increase, a periodic $d$-dependence is expected to disappear. Similarly, a non-periodic variation of the optical parameters has been predicted for highly absorbing materials [5]. Thus, the above trend is a characteristic of irradiation with Mid-IR pulses.

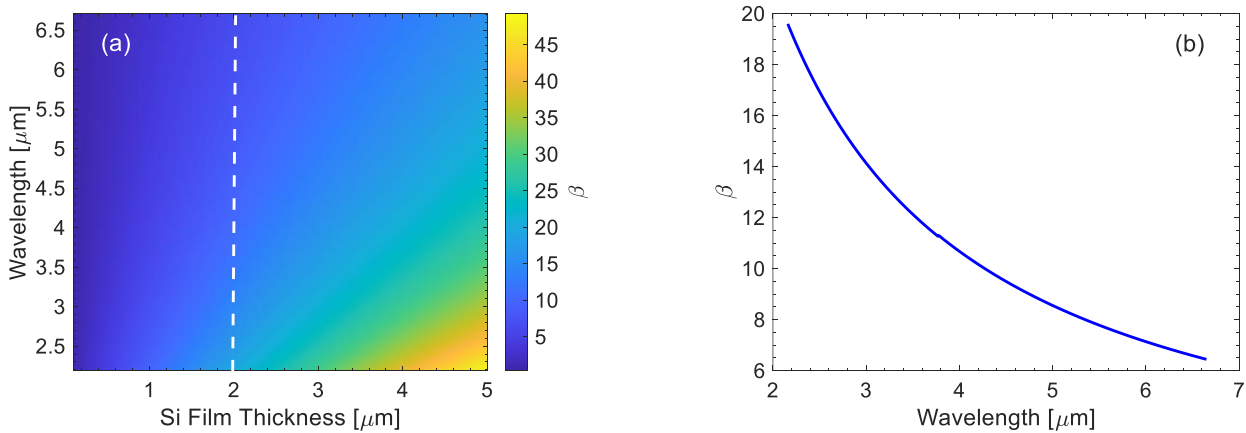

**Fig.SM4:** (a) Parameter $\beta$ as a function of $d$ and the laser wavelength, (b) $\beta$ along the *white cross line* in (a) (at $d=2$ μm).

## 4. Ultrafast dynamics and Optical response of Si/SiO$_2$ at $\lambda_L = 800$ nm and $\lambda_L = 3.2$ μm

An investigation of the ultrafast dynamics, optical and thermal response of the Si/SiO$_2$ has been performed to illustrate the differences arising at irradiation with $\lambda_L = 800$ nm and $\lambda_L = 3.2$ μm. To describe the behavior of Si/SiO$_2$ the following set equations (SM.3-4) are used to describe the optical response and ultrafast dynamics in the two materials:

### A. Silicon

#### i. $\lambda_L = 800$ nm

The carrier dynamics is provided by the following expression:

$$
\begin{aligned}
\frac{dN_e^{(Si)}}{dt} &= \frac{\alpha_{SPA}}{\hbar\omega_L}I + \frac{\beta_{TPA}}{2\hbar\omega_L}I^2 - \gamma\left(N_e^{(Si)}\right)^3 + \theta N_e^{(Si)} - \vec{\nabla}\cdot\vec{J} \\
\frac{dI}{dz} &= -(\alpha_{FCA} + \alpha_{SPA})I - \beta_{TPA}I^2 \\
I &= (1 - R - T)\frac{2\sqrt{\ln 2}}{\sqrt{\pi}\tau_p}Fe^{-4\ln 2\left(\frac{t-3\tau_p}{\tau_p}\right)^2}
\end{aligned}
\tag{SM.3}
$$

#### ii. $\lambda_L = 3.2$ μm

$$
\begin{aligned}
\frac{dN_e^{(Si)}}{dt} &= \frac{\beta_{TPA}}{2\hbar\omega_L}I^2 + \frac{\gamma_{TPA}}{3\hbar\omega_L}I^3 - \gamma\left(N_e^{(Si)}\right)^3 + \theta N_e^{(Si)} - \vec{\nabla}\cdot\vec{J} \\
\frac{dI}{dz} &= -\alpha_{FCA}I - \beta_{TPA}I^2 - \gamma_{TPA}I^3 \\
I &= (1 - R - T)\frac{2\sqrt{\ln 2}}{\sqrt{\pi}\tau_p}Fe^{-4\ln 2\left(\frac{t-3\tau_p}{\tau_p}\right)^2}
\end{aligned}
\tag{SM.4}
$$

In the above expressions, $\alpha_{SPA}, \beta_{TPA}, \gamma_{TPA}, \alpha_{FCA}$ stand for the single, two-photon, three-photon and free carrier absorption, respectively, $\vec{J}$ is the carrier current density and $R$ and $T$ correspond to the reflectivity and transmittance to the substrate, respectively; thus, $(1 - R - T)$ represents the energy absorptivity from the top layer. Simulation results show that at $\lambda_L = 3.2$ μm, $1 - R - T = 0$ which indicates that no energy is absorbed from Si. Furthermore, $N_e^{(Si)}$ correspond to the carrier density inside Si while $\theta$ stands for the impact ionisation coefficient. For a more detailed description of the ultrafast dynamics, see Ref. [3]. It is noted that only single, two-photon and three-photon absorption processes were considered and this might be confusing if longer laser wavelengths (i.e. smaller photon energies but within the Mid-IR spectral region) are used. Although multi-photon processes are required in that case, the fact that the absorptivity of energy from Si is zero yields no excitation of carriers and therefore, practically, $N_e^{(Si)} = 0$ at all Mid-IR pulses.

It is noted that in contrast to the coefficient '1-$R$' that has been taken in previous investigations (for example in Ref. [3]) to illustrate the percentage of the laser energy absorbed by Si (both for $\lambda_L = 800$ nm and $\lambda_L = 3.2$ μm), the coefficient '$(1 - R - T)$' is used in this study. This is due to the fact that in the current study, a double layered complex is used (and not a bulk material as in the former case) and therefore, some part of the energy is expected to be transmitted into the substrate.

### B. Fused Silica

On the other hand, the ultrafast dynamics and excited carrier evolution in fused silica are described for both $\lambda_L = 800$ nm and $\lambda_L = 3.2$ μm from Eq.2 (main manuscript) and Eq.SM.5.

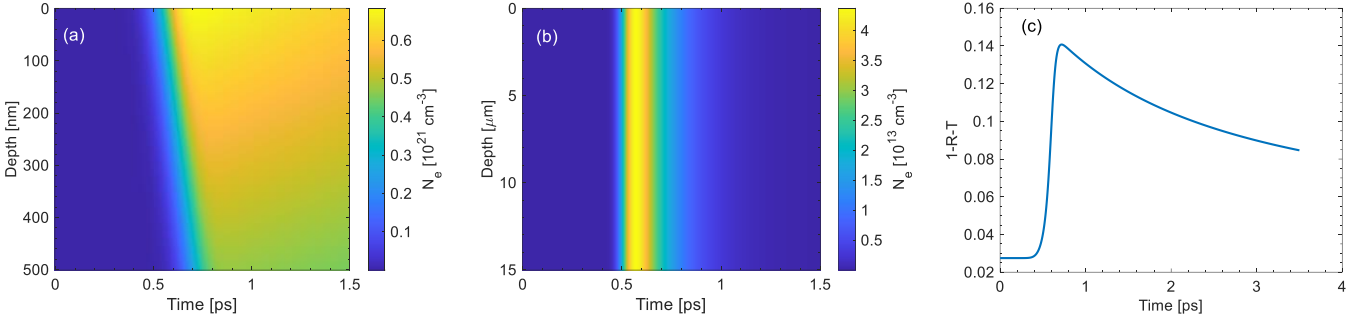

**Fig.SM5:** (a) Electron density evolution as a function of time and depth in Si, (b) Electron density evolution as a function of time and depth in SiO₂, (c) Absorptivity from Si (i.e. 1-*R*-*T*). (*F*=1 J/cm², *d*=500 nm, $\lambda_L = 800$ nm).

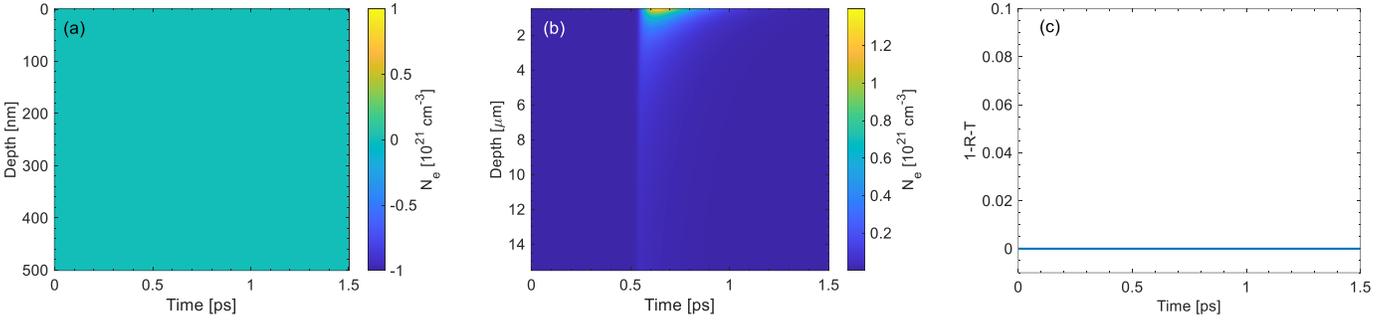

**Fig.SM6:** (a) Electron density evolution as a function of time and depth in Si, (b) Electron density evolution as a function of time and depth in SiO₂, (c) Absorptivity from Si (i.e. 1-*R*-*T*). (*F*=3 J/cm², *d*=500 nm, $\lambda_L = 3200$ nm).

Simulation results for the ultrafast dynamics in a double layered complex are illustrated in Fig.SM5 and Fig.SM6 for $\lambda_L = 800$ nm and $\lambda_L = 3.2$ μm, respectively. Furthermore, it is shown that while laser energy is absorbed from Si for $\lambda_L = 800$ nm, there is *no energy* absorption from Si for $\lambda_L = 3200$ nm (similar conclusions hold at other Mid-IR wavelengths as Si is transparent at Mid-IR). Without loss of generality, the simulations results illustrated in Figs.SM5-6 are depicted for *d*=500 nm at both wavelengths. Similar conclusions hold for other values of *d* (results are not shown). It is noted that simulations for $\lambda_L = 800$ nm were performed at *F*=1 J/cm² (to demonstrate the large number of excited carriers in Si at this fluence). Results show an energy absorption (and carrier excitation) in the Si film at $\lambda_L = 800$ nm while excited carriers are also created in SiO₂ due to a transmitted and absorbed amount of laser energy into SiO₂ (Fig.SM5). The energy absorptivity (1-*R*-*T*) from a 500 nm Si coating at 800 nm and 3200 nm are illustrated in Fig.SM5c and Fig.SM6c, respectively.

## 5. Model of Light propagation

In Fig.2b in the main manuscript, the spatio-temporal distribution of the excited electrons is illustrated; the depicted excited electron distribution results from the impact of the laser intensity *I* which is transmitted into the fused silica (as discussed in the previous Section, no energy absorption from Si is predicted). In our simulations, an attenuation along the volume of the local laser intensity due to the photoionisation and inverse bremsstrahlung (Free Carrier) absorption is considered that is provided in Eq.SM.5

$$\frac{dI}{dz} = N_{ph}^{(1)} \hbar \omega_L \frac{N_V - N_e}{N_V} W_{PI}^{(1)} + N_{ph}^{(2)} \hbar \omega_L \frac{N_{STE}}{N_V} W_{PI}^{(2)} - \alpha(N_e) I$$

$$I = A \frac{2\sqrt{ln2}}{\sqrt{\pi}\tau_p} F e^{-4ln2\left(\frac{t-3\tau_p}{\tau_p}\right)^2}$$

(SM.5)

where $N_{ph}^{(i)}$ corresponds to the minimum number of photons necessary to be absorbed by an electron that is in the valence band (*i*=1) or the band where the STE states reside (*i*=2) to overcome the relevant energy gap and reach the conduction band. On the other hand, *F* and $\omega_L$ stands for the laser fluence and frequency, respectively and *A* is the absorptivity from the substrate (SiO₂). Finally, the carrier density dependent parameter $\alpha$ corresponds to the free carrier absorption coefficient. More details can be found in Ref. [4].

## 6. Numerical Model

The theoretical framework that is used in this physical problem consists of: (i) a module that describes the laser energy distribution (i.e. following reflection, transmission and absorption and propagation), (ii) ultrafast dynamics, (iii) thermal effects (i.e. electron and lattice temperatures evolution) in the double layered complex. The model is aimed to derive, predominantly, these multiscale processes assuming Mid-IR pulses while, for the sake of comparison (see Section 4, above), a relevant analysis can be performed for pulses at lower wavelengths. A detailed description of the aforementioned processes following irradiation with fs Mid-IR pulses are provided in previous reports for Si [3] and SiO₂ [4], respectively.

An iterative Crank-Nicolson scheme based on a finite-difference method [5] has been used to solve the differential equations that describe ultrafast dynamics and thermal response [3, 4] and a multireflection technique is employed to determine the optical parameters [5] in the double layered material. Assuming the lack of absorption from the upper film (Si) at Mid-IR pulses, no carrier excitation occurs on Si and, therefore, we can consider that $N_e^{(Si)}$=0. By contrast, as the substrate absorbs the energy that is transmitted into SiO₂, electron excitation (and electron temperature rise) is only considered in SiO₂. The evolution of the electron and lattice temperature inside SiO₂ are calculated by a numerical scheme described in Ref. [4] which is based on the employment of a Two Temperature Model. A small (and insignificant) variation of the lattice temperature on Si, mostly, at the back end is calculated through the boundary conditions considered on the interface between the top layer and the substrate: $k_L^{(1)} \frac{\partial T_L^{(1)}}{\partial z} = k_L^{(2)} \frac{\partial T_L^{(2)}}{\partial z}$, where $k_L^{(1)}, k_L^{(2)}$ stands for the lattice heat conductivity in Si and SiO₂, respectively, while $T_L^{(1)}, T_L^{(2)}$ correspond to the lattice temperature in Si and SiO₂, respectively [5]. Finally, it is assumed that there is no carrier diffusion from Si into SiO₂. Thus, the boundary condition $\frac{\partial N_e^{(Si)}}{\partial z} = \frac{\partial N_e}{\partial z}$ =0 is considered on the interface. As noted above, for Mid-IR pulses, only $\frac{\partial N_e}{\partial z}$=0 (and $N_e^{(Si)}$ = 0) is used while for $\lambda_L$ = 800 nm, the expression $\frac{\partial N_e^{(Si)}}{\partial z} = \frac{\partial N_e}{\partial z}$ =0 is used.

## 7. Damage Threshold criterion: Thermal *vs* OBT criterion

In regard to the selection criterion of the damage threshold, it is noted that the condition $N_e > N_e^{cr}$ used in this work is a *necessary* but *not a sufficient* condition to produce damage of the dielectric material. In previous reports, it was proposed ([4], [6], [7-10]) that an energetic/thermal criterion (i.e. the lattice temperature exceeds the melting point of fused silica, $T_m$=1988 K) yields a more accurate estimation of the damage threshold. Nevertheless, in some previous studies, it was shown that the OBT-based criterion can still provide an approximate evaluation of the damage threshold [11-13]. Similarly, in a recent report where an analysis of irradiation of SiO₂ with Mid-IR pulses was presented [14], it was shown that there is a good agreement of the theoretical predictions (based on an OBT criterion) with previous experimental results [15]. Certainly, a more detailed investigation (experimental and theoretical) is required to determine which criterion is more accurate.

## 8. Photoinisation processes

As described in previous reports ([4,14]), photoinisation processes might be due to multiphoton/tunnelling ionization or a combination of multiphoton and tunnelling ionization. The value of the Keldysh parameter $\gamma$ can be used to determine which ionization process dominates. According to our previous conclusions Refs [4,14], we have assigned $\gamma<<1$ (or $\gamma>>1$) to define the regimes at which tunnelling ionization (or multiphoton ionization processes) dominate the photoionization processes. By contrast,
**Fig.SM7:** Keldysh parameter $\gamma$ during laser irradiation ($F$=3 J/cm², $d$=500 nm, $\lambda_L$ = 3200 nm).

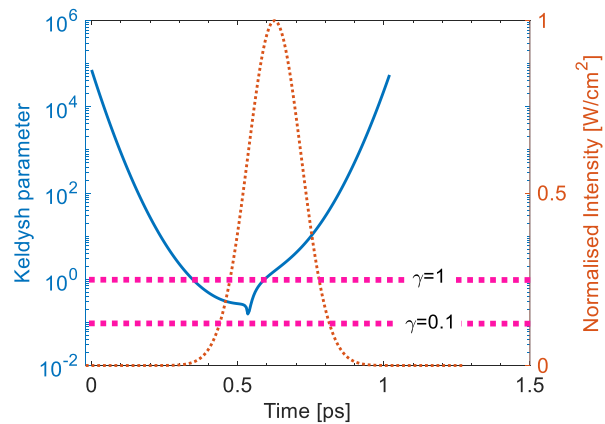

a combination of the two processes take place for intermediate values of $\gamma$. As the refractive index, transmissivity and laser intensity changes during irradiation, photoinisation rate varies. The dashed line (in *pink*) illustrates the border line for the various regimes.

Although results are shown here for a specific case ($F$=3 J/cm$^2$, $d$=500 nm, $\lambda_L = 3200$ nm), similar conclusions can be deduced in different conditions.

## 9. Damage Threshold

A parametric study has, also, been conducted to correlate the maximum carrier densities produced on the dielectric material due to electron excitation with the coating thickness (Fig. SM8a). Simulation results allow the estimation of the fluence threshold $F_{thr}$ at which the (maximum) value of $N_e$ exceeds a critical value $N_e^{cr}$ (i.e. $N_e^{cr} \equiv 4\pi^2 c^2 m_e \varepsilon_0/(\lambda_L^2 e^2)$) that is, usually, coined as the optical breakdown threshold (OBT). In particularly, $N_e^{cr} = 1.09 \times 10^{20}$ cm$^{-3}$ at $\lambda_L = 3.2$ μm. The capacity of specific values of the fluence to generate sufficient electron densities that excited SiO$_2$ above the OBT is illustrated in Fig. SM8a. More specifically, simulations have been conducted which show that the thickness of the coating influences the maximum carrier density $N_e^{max}$ on the surface of SiO$_2$ (the *red* dotted line in Fig. SM8a indicates the density of the produced electrons at $F$=3 J/cm$^2$). Furthermore, a correlation of the damage threshold with the coating thickness has been produced and theoretical results are illustrated in Fig. SM8b. Predictions yield values for $F_{thr}$ in the range between 2.0 J/cm$^2$ and 4.2 J/cm$^2$. Calculations show that for all Si film thicknesses, the presence of the coating induces damage of SiO$_2$ at higher fluences than in the absence of the Si film. The latter occurs at ~1.65 J/cm$^2$ (*red* dashed line in Fig. SM8b) [14]. To interpret this trend, it is important to evaluate the role of the optical parameters and the level of the absorbed energy from the substrate. As demonstrated in this work, the initial reflectivity appears to influence substantially the number of the generated carriers. For all Si thicknesses, the resulting reflectivity is always non-vanishing and larger than the reflectivity of SiO$_2$ which indicates that a higher fluence is required to damage the substrate compared to a bare SiO$_2$.

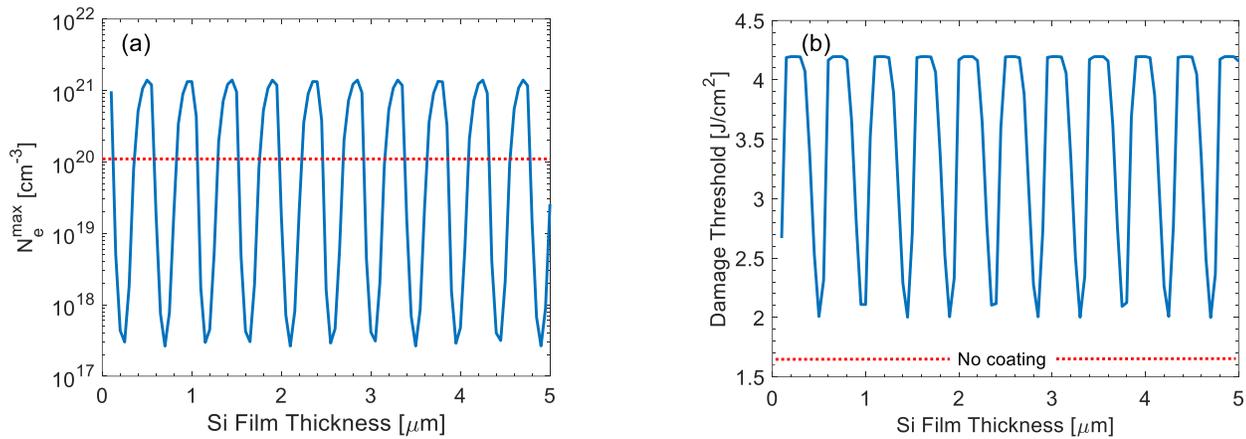

**Fig.SM8:** (a) Maximum electron density on the surface of SiO$_2$ at various $d$ ($F$=3 J/cm$^2$) (*red* dotted line indicates OBT). (b) Damage threshold on SiO$_2$ surface as a function of $d$ (*red* dotted line indicates the damage threshold for bulk SiO$_2$ in absence of coating that occurs at $F$=1.65 J/cm$^2$ [14]).